\documentclass[aps,showpacs,prb,twocolumn,superscriptaddress]{revtex4-1}%
\usepackage{indentfirst}
\usepackage{multirow}
\usepackage{color}
\usepackage{amsfonts}
\usepackage{amsmath}
\usepackage{mathrsfs}
\usepackage{amssymb}
\usepackage{wasysym}
\usepackage{graphicx}
\usepackage{CJK}
\usepackage{url}

\newcommand{\sumnear}{\mathop{\sum}_{\langle i j \rangle}}
\begin{document}
%
\title{High-Order Symbolic Strong-Coupling Expansion for the Bose-Hubbard Model}
\author{Tao Wang (ÍôÌÎ)}
\affiliation{School of Science, Wuhan Institute of Technology, 438000 Wuhan, China}
\affiliation{Physics Department and Research Center OPTIMAS,
Technical University of Kaiserslautern, 67663 Kaiserslautern, Germany}
\author{Xue-Feng Zhang (ÕÅÑ§·æ)}
\thanks{Corresponding author: {\tt zhangxf@cqu.edu.cn}}
\affiliation{Department of Physics, Chongqing University, Chongqing 401331,
	People's Republic of China} 
\affiliation{Physics Department and Research Center OPTIMAS, Technical University of Kaiserslautern, 67663
Kaiserslautern, Germany}
\affiliation{Max-Planck-Institut f\"{u}r Physik komplexer Systeme, N\"{o}thnitzer Strasse 38, 01187 Dresden, Germany}
\author{Chun-Feng Hou}
\affiliation{Department of Physics, Harbin Institute of
Technology, Harbin 150001, China}
\author{Sebastian Eggert}
\affiliation{Physics Department and Research Center OPTIMAS,
Technical University of Kaiserslautern, 67663 Kaiserslautern, Germany}
\author{Axel Pelster}
\affiliation{Physics Department and Research Center OPTIMAS,
Technical University of Kaiserslautern, 67663 Kaiserslautern, Germany}

\begin{abstract}
Combining the process-chain method with a symbolized evaluation we work out in detail a high-order symbolic strong-coupling expansion (HSSCE) for determining the quantum phase boundaries
between the Mott insulator and the superfluid phase of the Bose-Hubbard model for different fillings in hypercubic lattices of
different dimensions. With a subsequent Pad{\'e} approximation we achieve for the quantum phase boundaries a high accuracy, which is comparable to high-precision quantum Monte-Carlo
simulations, and show that all the Mott lobes can be rescaled to a single one. As a further cross-check, we find that the HSSCE results coincide with
a hopping expansion of the quantum phase boundaries, which follow from the effective potential Landau theory (EPLT).
\end{abstract}
\maketitle

%
\section{Introduction}
In recent decades, strongly correlated systems play a crucial
role in condensed matter physics. For high-temperature
superconductors,\cite{sc} the on-site interaction between fermions
becomes more dominant than the hopping processes.
Thus, considering the hopping term as a perturbation, the Hubbard
model can be reduced to the $t-J$ model or the Heisenberg model
at half filling. The bosonic counterpart is the Bose-Hubbard model,\cite{bh} 
which has  extensively been studied theoretically and can be realized 
experimentally using a gas of bosonic atoms in optical lattices.\cite{bh2,bh3} 
By reducing the tunneling processes via a deeper lattice potential or
by using the Fesh\-bach resonance technique in order to increase the interaction, those lattice systems can be tuned such that a quantum phase transition
from a  superfluid to a Mott insulator phase can be observed.\cite{bh2,bh3}
But also magnetic atoms,\cite{erbium} dipolar molecules,\cite{molecules} or
Rydberg atoms \cite{Rydberg} can be loaded into an
optical lattice, so that the strong long-range and anisotropic dipolar interaction 
plays a major role. In contrast to the weakly interacting
case, strongly dipolar correlated systems exhibit many exotic phases, such
as supersolid,\cite{ss1,ss2,ss3,triangular} superradiant solid,\cite{zhang2} or other topological phases.\cite{zhang3}

The Bose-Hubbard model,\cite{bh} which describes the quantum phase transition between the Mott insulator to the superfluid phase, is defined by the Hamiltonian
\begin{eqnarray}
\hspace*{-4mm}\hat{H}=-t\sumnear ( \hat{b}_{i}^{\dag}\hat{b}_{j}+\hat{b}_{j}^{\dag}\hat{b}_{i} ) +
\mathop{\sum}_i \left[ \frac{U}{2} \hat{n}_{i}(\hat{n}_{i}-1) -\mu \hat{n}_i \right],
\label{eqHam}
\end{eqnarray}
where $\langle i j \rangle$ represents a sum over 
nearest-neighbor sites, $t$ denotes the hopping matrix element, $\hat{b}_i^{\dag}(\hat{b}_i)$ creates (annihilates) a
particle on site $i$, $\hat{n}_{i}=\hat{b}_{i}^{\dag}\hat{b}_{i}$ abbreviates the number operator, 
$U$ stands for the on-site repulsion, and $\mu$ is the
chemical potential. 
The quantum phase transition was first established by using mean-field theory,\cite{bh}
but it is possible to obtain more accurate results by using a strong-coupling expansion (SCE) method, which 
was proposed by Freericks and Monien some time ago.\cite{Monien}
The strong-coupling ground state is given in the particle number
representation, while the hopping term is treated as a perturbation, 
so that the energy of both the Mott insulator and a single particle
(or hole) excited state can be calculated perturbatively. 
By equating the respective energies,
the critical line between the Mott insulator and the superfluid phase can be deduced. 
In comparison with the mean-field
approach,\cite{bh} this strong-coupling expansion method shows a higher accuracy 
for lower spatial dimensions, especially after
an extrapolation to higher orders. Therefore, SCE has been used successfully to study
the Bose-glass phase
in the superlattice,\cite{sc1} two-species bosons loaded into $d$-dimensional hypercubic
optical lattices,\cite{sc2,sc2b} and
the supersolid-solid quantum phase transition.\cite{zhang4} In particular, 
the strong-coupling expansion method has turned out to be
efficient for the second-order transition from an
incompressible to a compressible phase. However, the
calculation effort turns out to increase with the order and filling in form of a power law, so analytic expressions from
SCE are usually limited up to the fourth order.  

In order to obtain higher orders  than the $10$th order, Eckardt {et al.}~developed a computer assisted 
process-chain algorithm (PCA) \cite{eckardt,martin}
based on Kato's formulation of the
perturbation calculus.\cite{kato}
Using the PCA, high-precision results were obtained for both the ground-state energy and the
correlation function within the Mott insulator.\cite{martin} Furthermore, by
implementing PCA for the effective potential Landau
theory (EPLT),\cite{santos,tao2} the critical line of the quantum phase transition
from the Mott insulator to the superfluid phase was determined with 
high precision \cite{martin,ying} so that even the corresponding critical exponents
can be extracted.\cite{hinrichs}
In order to deal with degenerate states, such as particle-hole excitations, the PCA 
must be modified, as was shown by Heil and Linden for a one-dimensional (1D) system.\cite{Heil}
An alternative SCE approach by Elstner and Monien also obtained 
high-order results which can be applied to low dimensions and low filling.\cite{Elstner} However, a general analytic expression for quantum phase boundaries of different fillings,
orders, and dimensions is still lacking. They may help, for instance, to hint at a scaling relation between different fillings.\cite{Teichmann}
In addition, it is still unclear how SCE and EPLT are related although both methods deal with a perturbative hopping expansion.

In this paper, we work out a high-order symbolic SCE algorithm (HSSCE) and combine it with
PCA for degenerate states in order to obtain the quantum
phase boundary of the Bose-Hubbard model. To this end we briefly review in Sec.~II the strong-coupling
expansion, for which we propose an efficient method and list the {\it general analytic} strong-coupling series
up to eighth order on chain (1D), square (2D), and cubic (3D) lattices for arbitrary filling $n$.
Then the respective quantum phase boundaries are obtained in Sec.~III  in the thermodynamic limit by applying the Pad{\'e} resummation method to the strong-coupling series and
by comparing the results with those from  numerical high-precision calculations.
In Sec.~IV, based on the analytic expression of the critical line for arbitrary filling, we rescale the Mott lobes to the infinite filling Mott lobe. Afterwards, in Sec.~V,
we show that the HSSCE results coincide with a hopping expansion of the high-order effective potential Landau theory (HEPLT). Finally, we draw our conclusions in Sec.~VI. In addition, in order to
assist writing a code, we work our algorithm in detail in Appendix A and also attach the Matlab code in the Supplemental Material.\cite{sm}

\section{Strong-Coupling Expansion for the Bose-Hubbard Model}\label{sec_sc}
The strong-coupling expansion method is based on treating effects of the hopping matrix element perturbatively.
It can be used to determine the second-order critical line of the incompressible phase, whose
melting is caused by a proliferation of particle or hole excitations.
Based on the work of Freericks and Monien,\cite{Monien} one determines at first
the unperturbed ground-state wave function of both the incompressible
state and the particle (hole) excited state.
Then one calculates the respective ground-state energies in terms of a hopping expansion by applying
non-degenerate and degenerate perturbation theory, respectively. At last, by
comparing the resulting ground-state energies, the corresponding critical line is deduced.
\begin{figure}[t]
\includegraphics[width=0.5\textwidth]{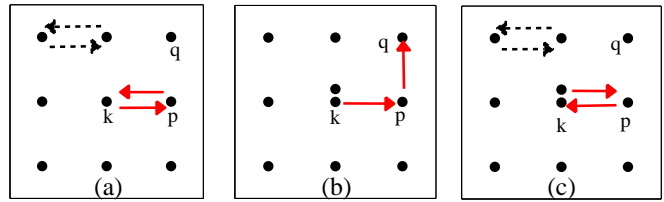}
\caption{Second-order arrow diagrams for (a) the ground state of a Mott insulator
of filling one, and also for (b) open and (c) closed arrow
diagrams for the case of adding one particle. As all arrow diagrams
irrelevant to site $k$, indicated by black dashed arrows, cancel each other, only
the arrow diagrams relevant to site $k$, highlighted by red solid arrows, need to be
considered.}\label{example}
\end{figure}

Let us consider the second-order quantum phase transition of a Mott insulator as a concrete 
example, which is described by the Bose-Hubbard model (\ref{eqHam}).
The dominant part is provided by the on-site repulsive interaction together with the
chemical potential, i.e. $\hat{H}_0=\mathop{\sum}_i \left[ U
\hat{n}_{i}(\hat{n}_{i}-1) / 2 -\mu \hat{n}_i \right]$, whereas the
perturbative part is the hopping term
$\hat{H}'=-t\sumnear(\hat{b}_{i}^{\dag}\hat{b}_{j}+\hat{b}_{j}^{\dag}\hat{b}_{i})$.
When the tunnel matrix element $t$ vanishes, the ground
state of the Mott insulator with filling $n$ is non-degenerate and uniquely given by
$|\psi^{(0)}_M \rangle =\mathop{\prod}_i(\hat{b}_{i}^{\dag})^n|0\rangle/\sqrt{n!}$.
In contrast to that, the particle excited state $\hat{b}_{j}^{\dag} |\psi^{(0)}_M \rangle$
is degenerate for $t=0$, since no matter at which site the additional
particle is located, the corresponding ground-state energies coincide.

The ground-state energy of the Mott insulator reads in zeroth order of the tunnel matrix element
$E^{(0)}_M= N [Un(n-1)/2-\mu n]$, where $N$ denotes the number of lattice sites. The 
first-order ground-state energy correction turns out to be zero, so 
the first non-vanishing correction is of second-order and follows from
$E^{(2)}_M=\mathop{\sum}_{i\ne0}\langle \psi^{(0)}_M |H'|e_i\rangle\langle
e_i|H'| \psi^{(0)}_M \rangle/(E^{(0)}_M-E_i)$, where $|e_i\rangle$ denotes an excited
state with energy $E_i$. According to Fig.~\ref{example} (a), the
second-order processes correspond to the case that each particle is hopping to the nearest
neighbor site and then back. Thus, the second-order perturbation
energy yields $E^{(2)}_M=-z(n+1)Nt^2/U$ where $z=2d$ stands for
the coordination number of a $d$ dimensional hypercubic lattice.
Then the ground-state energy of the Mott insulator up to second order in the tunnel matrix element
is given by $E_{M}=E^{(0)}_M-z(n+1)Nt^2/U$.

For the particle excited state, the zeroth-order ground-state energy is given by $E^{(0)}_p=E^{(0)}_M-\mu+nU$. The general form of the ground
state wave function is given by the superposition
$|\psi_p^{(0)} \rangle =\mathop{\sum}_ja_j \hat{b}_{j}^{\dag} |\psi^{(0)}_M \rangle$ with the
normalization constraint $\mathop{\sum}_j|a_j|^2=1$. The first-order ground-state energy is then determined via
$E^{(1)}_p=\langle \psi_p^{(0)} |\hat{H}'| \psi_p^{(0)} \rangle$. In order to
minimize $E^{(1)}_p$, we need to diagonalize the matrix
$M_{i,j}= \langle \psi^{(0)}_M | \hat{b}_{i} \hat{H}'\hat{b}_{j}^{\dag} | \psi^{(0)}_M \rangle$ and take the
lowest eigenvalue as the resulting first-order ground-state energy. Thus, in other words, 
first-order perturbation lifts the degeneracy. Based on the underlying
translational symmetry of the ground state, we obtain
the first-order ground-state energy $E^{(1)}_p=-z(n+1)t$ and the non-degenerate ground
state turns out to be $| \psi_p^{(0)} \rangle =\mathop{\sum}_j\hat{b}_{j}^{\dag} | \psi_M^{(0)} \rangle /\sqrt{Nn}$. Now we turn to second-order processes,
where Fig.~\ref{example}(b) and (c) depict via red solid arrow diagrams the two possible types, which occur in the presence of
one additional particle. The
first diagram is an open diagram, which is characterized by $|211\rangle\rightarrow|202\rangle\rightarrow|112\rangle$
with $|n_kn_pn_q\rangle$ denoting the occupation of the respective sites $k$, $p$, $q$
and has the corresponding energy $-(n+1)t^2/U$.
The second one is the closed diagram $|21\rangle\rightarrow|30\rangle\rightarrow|21\rangle$
with $|n_kn_p\rangle$ representing the occupation of the sites $k$, $p$
and the related energy $-(n+2)t^2/(2U)$.  Note that the open
diagram Fig.~\ref{example}(b) has the multiplicity $z(z-1)$, whereas the multiplicity of the closed one in Fig.~\ref{example}(c) is given by $z$. Besides that,
we have to take into account in Fig.~\ref{example}(b) and (c) that there are additional $z(N-2)$ closed diagrams, indicated by black dashed arrow diagrams, which are not
related to the additional particle. Thus, the second-order ground-state energy
results in $E_p^{(2)}=-[z(N-2)(n+1)+z(n+2)/2+z(z-1)(n+1)]t^2/U$ and the
total energy of the particle excited state is given by
$E_p=E^{(0)}_M-\mu+nU-z(n+1)t+E_p^{(2)}$. Finally, equating the Mott and the particle excited ground-state energy according to $E_M=E_p$, we get
the following critical line up to second order in the tunnel matrix element $t$: $\mu_p=nU-z(n+1)t-[z(n+2)/2+z(z-1)(n+1)-2z(n+1)]t^2/U$.

However, the number of arrow diagrams is of the order of the lattice size $N$, so this calculation is not practical in the thermodynamic
limit even with the process-chain approach. 
Therefore, we introduce here a novel strong-coupling method, which turns out to be more practical and efficient
as it is quite suitable for a computer implementation. From the above derivation of the second-order result for the critical line, we read off that perturbative
calculations follow more easily from neglecting all energy contributions, which are not related to the
site $k$, where the additional particle is located, as all those contributions irrelevant to site $k$ in $E_M$ and
$E_p$ cancel each other at the end. Thus, in order to obtain the critical
line to higher orders in the tunnel matrix element most efficiently, we need to consider the following three steps:\\
(i) We calculate the energies of the
diagrams related to the site $k$ based on the ground state
$|\psi^{(0)}_M \rangle$, and name them \emph{energy corrections}. To this end
we only need to find all closed arrow diagrams related to site $k$ 
and calculate the contribution for each diagram to the non-degenerate ground-state energy.
In Fig.~\ref{example} (a), we have $z$ hopping processes, where a
particle at site $k$ hops to a nearest neighbor site and then back, as well as
$z$ hopping processes, where a particle at a neighbor site hops to
site $k$ and back, thus the second-order energy correction results is	
$-2z(n+1)t^2/U$.\\
(ii) We only determine the energy, which is related to the
additional particle, and name it \emph{strong-coupling (SC) 
energy}. To this end we need to find all arrow diagrams related to site $k$
and calculate the contribution of each diagram to the degenerate ground-state energy.
In Fig.~\ref{example}, the diagrams (b) and (c) have
the multiplicity $z$ and $z(z-1)$, respectively, thus the corresponding SC energies read 
$-z(n+2)t^2/(2U)$
and $-z(z-1)(n+1)t^2/U$.\\
(iii) Equating the energy of processes
(i) and (ii), we obtain the resulting SCE critical line. \\
Based on this strong-coupling expansion method, we only need to
consider a finite number of diagrams, combine their evaluation with the process-chain method,
and then obtain their perturbative value. Note that for the
case of the hole excitation, the calculations proceed similarly, except that the arrows point then in the direction, where the hole moves. 

In order to implement this method, we used an algorithm 
proposed by Heil and Linden,\cite{Heil} but in our case each step is realized in terms of a symbolic calculation.
Then, the representation of the coefficients in each order is an analytic
function of both the filling $n$ and the hopping amplitude $t$. In order to make the implementation of the algorithm more explicit, we explain
the details in Appendix A, and make the corresponding Matlab
code available in the Supplemental Material.\cite{sm}
With this we have determined the critical lines of the quantum phase diagram of the Bose-Hubbard model for a general Mott lobe $n$, which were obtained with symbolic calculations up to the
8th order. The strong-coupling results for both the upper and the lower
critical line of the Mott insulator with respect to both
particle and hole excitations are defined according to
\begin{equation}
\label{particle}
\frac{\mu_{\rm p}}{U}=1 -\mathop{\sum}_{i=1}^{\infty}\beta_{\rm u}^{(i)} \left( \frac{t}{U} \right)^i
\end{equation}
and 
\begin{equation}
\label{hole}
\frac{\mu_{\rm h}}{U}=\mathop{\sum}_{i=1}^{\infty} \beta_{\rm d}^{(i)} \left( \frac{t}{U} \right)^i \, ,
\end{equation}
respectively. The analytic expressions with coefficients up to 8th order 
are given in Appendix B.

\begin{figure}[t]%
	\includegraphics[width=0.49\textwidth]{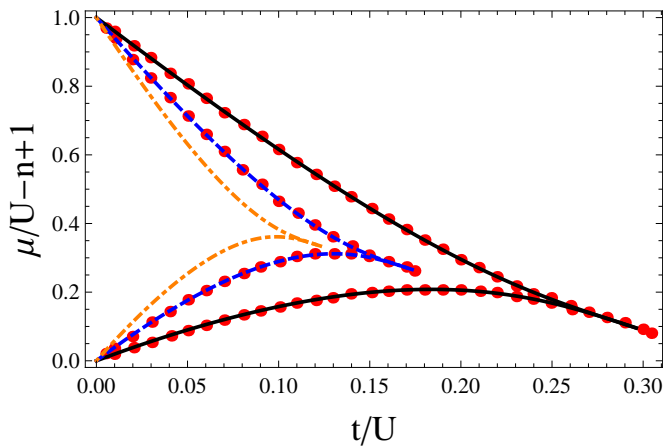}
	\caption{Quantum phase diagrams of Bose-Hubbard chain for filling $n=1$ (black solid line), $n=2$ (blue dashed line), and $n=3$ (orange dot-dashed line).
          The red dots are DMRG results for filling $n=1$ and $n=2$.\cite{Gebhard}}\label{1dphase}
\end{figure}
\section{Quantum Phase diagram for Bose-Hubbard model}\label{sec_re}
In order to reconstruct from such perturbative results the whole
quantum phase diagram in the thermodynamic limit, we need to know the scaling behaviour of the quantum phase transition. According to previous works,\cite{bh,Sachdev}
the Mott-superfluid quantum phase transition of the Bose-Hubbard model with dimension $d\ge2$
belongs to the $d+1$ dimensional $XY$ universality class.  But in one dimension
the quantum phase transition turns out to be of the Berezinsky-Kosterlitz-Thouless (BKT) type, which has to be treated separately.
\subsection{BKT Universality Class}
In the one-dimensional case the quantum phase boundary can be rewritten as $\mu_{p(h)}=B(t) \pm \Delta(t)/2$, where the strong-coupling series for both the mean energy $B(t)=(\mu_{p}+\mu_h)/2$
and the energy gap $\Delta(t)=\mu_{p}-\mu_h$ follow from
from Eqs.~(\ref{1dup}) and (\ref{1ddown}), respectively. On the other hand we know for
the universality class of Berezinsky-Kosterlitz-Thouless that the energy gap $\Delta(t)$ is characterized by a non-analytic behaviour slightly below the critical point $t_c$
according to \cite{Monien,Monien2}
\begin{equation}
\Delta(t)=A\exp\left(-\frac{W}{\sqrt{t_c-t}}\right)\,, 
\label{gap1}
\end{equation}
with some coefficients $A$ and $W$, so we conclude $[\log\Delta(t)]^2 \propto (t_c-t)^{-1}$. Such a divergent behaviour could be recovered from the strong-coupling series
of $[\log\Delta(t)]^2$, which is available in terms of Eqs.~(\ref{1dup}) and (\ref{1ddown}) in Appendix B.
up to the $8$th order by applying the Pad{\'e} resummation method \cite{padebook}
\begin{equation}
[\log\Delta(t)]^2=\frac{{\displaystyle \sum_{m=0}^{m_{\rm max}}} a_{m}t^{m}}{1+{\displaystyle \sum_{n=1}^{n_{\rm max}}}b_{n}t^n}\, .
\label{pade}
\end{equation}
To this end the corresponding Taylor series of the left- and right-hand side of Eq.~(\ref{pade}) are used to obtain the respective coefficients $a_{m}$ and $b_{n}$ from fitting with the restriction
$n_{\rm max}+m_{\rm max}=8$.
The most natural way to achieve this is to choose $n_{\rm max}=m_{\rm max}=4$. Note that we also have to adopt a Pad{\'e} resummation similar to Eq.~(\ref{pade}) for the strong-coupling series of the mean energy $B(t)$  
in order to describe the re-entrance behaviour of the Mott lobe, which is typical for a phase transition of the  Berezinsky-Kosterlitz-Thouless type.
In this way we determine the whole quantum phase diagram for different fillings $n$. 
\begin{table}
  \begin{center}
		\begin{tabular}{|c|c|c|c|}
			\hline
			&\multicolumn{2}{c|}{ $~$HSSCE$~$ } &\multicolumn{1}{c|}{numerical } \\
			\cline{2-4}
			& $t_{c}/U$ & $\mu_{c}/U$ & $t_{c}/U$  \\
			\cline{2-4}
			&\multicolumn{3}{c|}{ $~$$d=1$$~$ } \\
			\hline
			$n=1$& $0.296 $ & $0.0956$ & $0.30(1)$   \\
			\hline
			$n=2$& $0.173$ & $1.2659$ &  $0.175(2)$  \\
			\hline
			$n=3$& $0.123$ & $2.3339$ &    \\
			\hline
			&\multicolumn{3}{c|}{ $~$$d=2$$~$ } \\
			\hline
			$n=1$ & 0.05989& 0.3705& 0.05974(3)  \\
			\hline
			$n=2$ & 0.03530 & 1.4238&  \\
			\hline
			$n=3$ &0.02509 &2.4459&   \\
			\hline
			&\multicolumn{3}{c|}{ $~$$d=3$$~$ } \\
			\hline
			$n=1$ & 0.03415 & 0.3929& 0.03408(2)   \\
			\hline
			$n=2$ & 0.02013 & 1.4369 & \\
			\hline
			$n=3$ &0.01431 &2.4552&    \\
			\hline
		\end{tabular}
		\caption{Critical hopping amplitude and 
			chemical potential at the lobe tip for different dimensions $d$ and filling numbers $n$. The HSSCE results are determined in combination with an $8$th order Pad{\'e} resummation. 
			The $1d$ numerical result stems from DMRG calculations,\cite{Gebhard,Kollath,Polkovnikov} while
			those for $2d$ and $3d$ are obtained from QMC simulations.\cite{sv1,sv2}}\label{critical point}
\end{center}\end{table}

As shown in Fig.~\ref{1dphase}, a quantitative comparison with DMRG calculations \cite{Gebhard} at low filling $n$ demonstrates that the quantum phase boundaries determined from HSSCE together with a
Pad{\'e} resummation reveal a high accuracy except from tiny deviations at the lobe tip. In particular, the critical hopping amplitude $t_{c}$ coincides
in our method with a real, positive simple pole of Eq.~(\ref{pade}), which turns out to be unique. Table~\ref{critical point} lists the critical hopping amplitude $t_{c}$ for
the first three fillings $n$. Our results combined with various DMRG calculations have established the consensus that $t_c / U \approx 0.3$ for $n=1$.
But note that
the results of DMRG calculations for the critical point $t_c$ have also relatively large uncertainties and 
slightly deviate depending on the chosen observables since the gap becomes so small and requires extremely large system sizes.
For filling $n=1$, initial estimates were given to be $t_c/U=0.277 \pm 0.01$ \cite{Monien2} and have now been improved to $t_c/U=0.3050 \pm 0.001$ from density density correlations \cite{Gebhard} (the red dot shown in Fig.~\ref{1dphase}),
$t_c/U=0.2980 \pm 0.005$ from the von-Neumann entropy,\cite{Kollath} and $t_c/U=0.3030 \pm 0.009$ from the energy gap.\cite{Polkovnikov}
In addition, for increasing fillings $n$, we observe that the re-entrance behaviour weakens and tends to disappear. 
This result is an immediate consequence of the particle-hole symmetry
which is recovered in the limit of infinite filling $n \rightarrow \infty$ irrespective of the spatial dimension.
Indeed, the Hamiltonian (\ref{eqHam}) does not change with the transformation $\hat{b}^{\dag}\rightarrow \hat{b}'$,
$\hat{b}\rightarrow \hat{b}'^{\dag}$, which implies for the filling $n\rightarrow n'+1\approx n'$. This is reflected in the strong-coupling results for
the quantum phase boundaries listed in Appendix B by the property that the coefficients of the highest filling 
in the upper and the lower branch have the same absolute value but a different sign. 

\begin{figure}[t]
\includegraphics[width=0.49\textwidth]{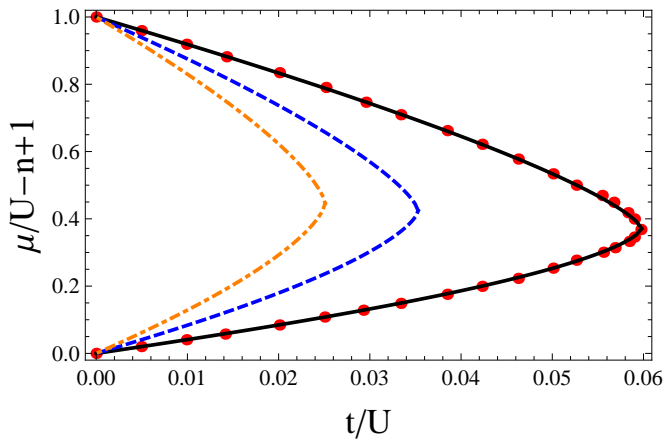}
\caption{Quantum phase diagram of $2$d bosonic lattice system for $n=1$ (black solid line), $n=2$ (blue dashed line), and $n=3$ (orange dot-dashed line). For
$n=1$ we also show QMC simulation results (red dot).\cite{sv2}}
\label{2dphase}
\end{figure}
\subsection{XY Universality Class}
\begin{figure}[t]
\includegraphics[width=0.485\textwidth]{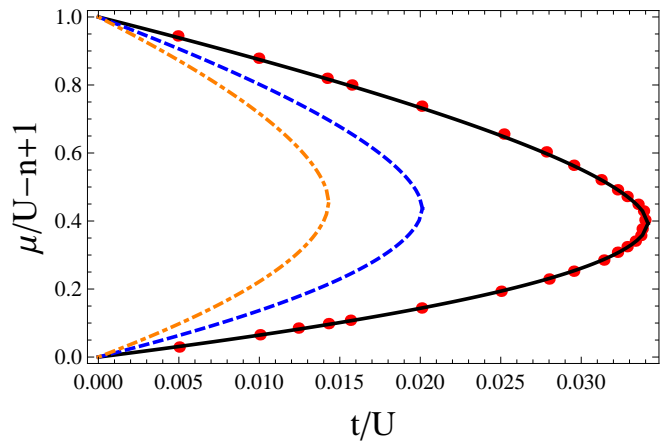}
\caption{Quantum phase diagram of $3$d bosonic lattice system for $n=1$ (black solid line), $n=2$ (blue dashed line), and $n=3$ (orange dot-dashed line).
For $n=1$ we also show QMC simulation results (red dots).\cite{sv1}} \label{3dphase}
\end{figure}

For higher dimensional systems $d\ge2$, the energy gap follows slightly below the critical point $t_c$
the scaling law $\Delta(t)=A(t)(t_c-t)^{z\nu}$.\cite{bh,Sachdev} Here $A(t)$ represents a regular function, whereas $z$ and $\nu$ denote critical exponents, which characterize the dynamics and
the correlation function of the respective universality class.\cite{Zinn-Justin,Kleinert}
Thus, also $\partial (\log \Delta(t) )/\partial t$ has a simple pole at $t_c$, which can be determined via the Pad{\'e} resummation method. The resulting quantum phase diagram for different
fillings $n$ for the dimensions $d=2$  and $d=3$ are shown in Figs.~\ref{2dphase} and \ref{3dphase}, respectively. We conclude that the analytically obtained
quantum phase boundaries match quite well with QMC simulation result at filling $n=1$ in both two and
three dimensions. The real, positive simple pole of the Pad{\'e} resummation, which turns out again to be unique,
provides the critical points $t_c$ for different filling numbers $n$,  where the lowest three values are listed in Table~\ref{critical point}. Moreover, the
corresponding residue yields a value for the critical exponent $z\nu$ listed in Table~\ref{critical exponent}. Our results reveal that $z\nu$ slightly depends on the filling number $n$, but still deviates from
the critical exponent of XY models.\cite{xy} Thus, higher order terms may need to be considered in a future work, in order to diminish that deviation.

\begin{table}\begin{center}
		\begin{tabular}{|c|c|c|c|c|c|}
			\hline
			& $n=1$ & $n=2 $ &$n=3$ & $n\rightarrow\infty$ & XY \\
			\hline
			$d=2$& $0.6965$ & $0.6951$ &  $0.6948$ &$0.6945$ &$0.6715$ \cite{xy} \\
			\hline
			$d=3$ & $0.5625$ & $0.5647$ & $0.5656$ &$0.5663$  &$0.5$ \\
			\hline
		\end{tabular}
		\caption{Critical exponents $z\nu$ of Mott-superfluid phase transitions in two and three dimensions.}\label{critical exponent}
\end{center}\end{table}

\section{Rescaling Properties of different filling}\label{sec_sp}

In this section we investigate systematically the filling-dependent rescaling properties of the Bose-Hubbard quantum phase diagram and 
restrict ourselves to the case of $2$d and the $3$d.
At first, we consider for different fillings $n$ the rescaling properties of the critical point $t_c(n)$, which represents the tip of the Mott lobe. To this end we follow Ref.~[\onlinecite{Teichmann}]
and use its second-order strong-coupling result in order to re-express the hopping amplitude of the lobe tip for large filling $n$
according to
\begin{eqnarray}
  \frac{t_c(d,n)}{U} &=&  \frac{2d-\sqrt{-10d+12d^{2}}}{(10-8d)d} \nonumber\\
  && \times \left[1-\frac{1}{2n}+O\left(\frac{1}{n^{2}}\right)\right]\,\frac{1}{n}\,.
\end{eqnarray}
Thus, in the limit of large filling, the effects of dimension $d$ and filling $n$ turn out to decouple. Assuming that such a decoupling property
also holds for higher strong-coupling orders, we perform for the critical hopping amplitude the generic ansatz
\begin{equation}
 \frac{t_c(d,n)}{U} =f_t(d) \, \frac{g_t(n)}{n} \, ,
  \label{tylor}
\end{equation}
where the filling-dependent factor represents a Taylor series in $1/n$: 
\begin{equation}
g_t(n)=1+\sum_{i=1}^{\infty}\frac{a_{i}}{n^{i}}\,.
\end{equation}
In the limit of infinite filling we then conclude
\begin{equation}
f_t(d)=\lim_{n \rightarrow \infty} n \frac{t_c(d,n)}{U} \,.
\end{equation}
This means that $f_t(d)$ can be calculated both for $d=2$ and $d=3$ from the term with highest filling number $n$ in each strong-coupling order for both the upper and the lower quantum phase boundary presented
in Appendix B. Note that, due to the particle-hole symmetry mentioned above, the coefficients of the highest filling 
in the upper and the lower branch have the same absolute value but a different sign. 

\begin{figure}[t]
\includegraphics[width=0.5\textwidth]{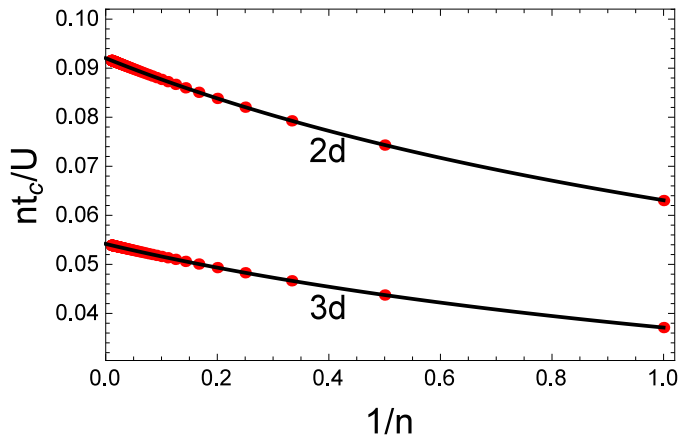}
\caption{Critical hopping amplitudes $t_c$ for filling numbers $n$ from $1$ to $100$ (red dots) compared with the value of $f_t(d)$ times the rescaling
  function $g_t(n)$ (black line) in two and three dimensions.}\label{scaling}
\end{figure}

After having obtained $f_t(d)$ for $d=2$ and $d=3$, we directly read off from Eq.~(\ref{tylor}) in each strong-coupling order the function 
$g_t(n)=n t_c(d,n)/[f_t(d) U] $ as a Taylor series in $1/n$. In order to approach the infinite-order case, we perform a Pad\'e resummation
and rewrite the filling-dependent function as
\begin{equation}
g_t(n)\approx \tilde{g}_t(n)=\frac{1+{\displaystyle \sum_{i=1}^{M} \frac{\alpha_{i}}{n^{i}} }}{1+{\displaystyle \sum_{j=1}^{M} \frac{\beta_{j}}{n^{j}}  }}\,.\label{scal}
\end{equation}
Here $M$ is an integer, which characterizes the order of the Pad\'e resummation. The resulting function $\tilde{g}_t(n)$
represents the rescaling function of the critical hopping amplitude. In Fig.~\ref{scaling} we have chosen $M=4$ and have used the critical
hopping amplitude at the tip of the first 100 Mott lobes to fit the scaling function. 
In addition it turns out that the rescaling  functions in two and three dimensions nearly coincide, which supports a posteriori the above assumption from Eq.~(\ref{tylor}) 
that dimension $d$ and filling $n$ decouple. 

Furthermore, one can also use a similar strategy in order to investigate the rescaling properties of the critical chemical potential $\mu_{c}(d,n)/U$ at the lobe tip. But then
the above mentioned particle-hole symmetry for infinite filling implies for the chemical
$\lim_{n \rightarrow \infty}\mu_p/U=1-\lim_{n \rightarrow \infty}\mu_h/U$, so the critical chemical potential reads
$\lim_{n \rightarrow \infty}\mu_c/U=1-\lim_{n \rightarrow \infty}\mu_c/U=1/2$.
Thus, the dimension dependent function is then given by $f_{\mu}(d)=\lim_{n \rightarrow \infty}\mu_c/U=1/2$, so the rescaling function $g_{\mu}(d,n)=2 \mu'_c(d,n)/U$
with $\mu_{c}'(d,n)/U=\mu_{c}(d,n)/U-n+1$ results as a Taylor series in $1/n$, for which we perform a Pad\'e resummation
\begin{equation}
g_{\mu}(d,n)\approx\tilde{g}_{\mu}(d,n)=\frac{1+{\displaystyle \sum_{i=1}^{M} \frac{\alpha'_{i}}{n^{i}} }}{1+{\displaystyle \sum_{j=1}^{M} \frac{\beta'_{j}}{n^{j}} }}  \, .
\end{equation}
Note that, in contrast to the critical hopping, the rescaling function $\tilde{g}_{\mu}(d,n)$ for the chemical potential turns out to have a residual dependence on the dimension $d$.

\begin{figure}[t]
\includegraphics[width=0.5\textwidth]{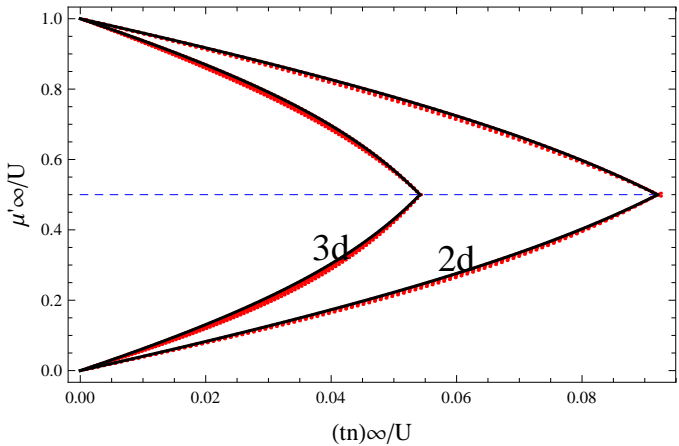}
\caption{Scaled Mott lobe obtained from different fillings $n$ (red dots) compared with infinite filling quantum phase diagram (black line). For both dimensions $d=2$ and $d=3$ the scaled Mott lobes deduced from filling
  number $n=1,2,3,4,10,100,10000$ are almost on top of each other.}\label{generalphase}
\end{figure}

By assuming that the entire critical lines have the same rescaling  functions at the tips of the Mott lobes, we can map all Mott lobes for different filling numbers to the infinite filling lobe as follows.
For each critical point $\{\mu'(d,n)/U , n t(d,n)/U\}$ in the lower branch of the Mott lobe, we define the rescaled value as $\{\mu'_{\infty}/U,(tn)_{\infty}/U \}=\{\mu'(d,n)/[\tilde{g}_{\mu}(n)U],t(d,n)n/[\tilde{g}_{t}(n)U]\}$
and, correspondingly,
for each critical point in the upper branch we rescale according to
$\{\mu'_{\infty},(tn)_{\infty}/U\}=\{1-[1-\mu'(d,n)/U]/[2-\tilde{g}_{\mu}(n)],t(d,n)n/[\tilde{g}_{t}(n)U]\}$. In Fig.~\ref{generalphase} we observe that all scaled Mott lobes, obtained
from different filling numbers $n$, deviate only slightly from the quantum phase boundary at infinite filling.

Finally, we comment upon why the rescaling property turns out to be more complicated for the one-dimensional system. Although 
we could also find a rescaling  function for the critical point, this rescaling function could not be used to map all the lobes to the infinite filling
lobe. Whereas a Mott lobe with finite filling reveals a re-entrance phenomenon due to the BKT quantum phase transition, this re-entrance phenomenon disappears in the limit of an infinite
filling due to the particle-hole symmetry between the upper and the lower phase boundary. Thus, in one dimension there does not exist
a universal rescaling function for all points of the quantum phase diagram. 

\section{Relation with high-order effective potential landau theory}\label{sec_cp}

As has already been explained in the introduction, both SCE \cite{Monien} and EPLT \cite{santos,tao2} represent two analytical perturbative methods for determining the quantum phase boundary of lattice
systems. Usually the accuracy of their results are only compared in lower orders. Thus, in order to allow for a 
comparison in higher orders, we have also investigated the $2$d Bose-Hubbard model with the
HEPLT method from Ref.~[\onlinecite{martin}] up to the $10$th order. Whereas the HSSCE method allows to determine the upper or the lower quantum phase boundary with one single calculation, 
HEPLT is more involved and needs independent calculations to obtain the critical hopping for each chemical potential. Note that, up to the same order, HSSCE turns out to be faster than HEPLT
because it has not to deal with additional source terms.

\begin{figure}[t]
	\includegraphics[width=0.5\textwidth]{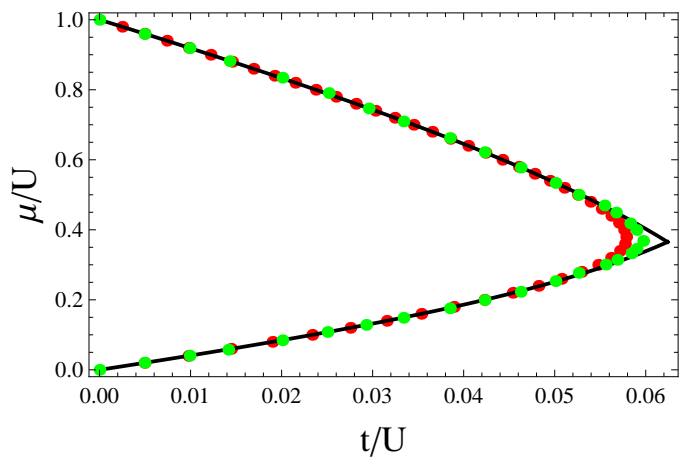}
	\caption{Comparison of quantum phase diagrams for $d=2$ via HSSCE (black line) and HEPLT (red dots)\cite{martin} in $10$th order with QMC results (green dots)\cite{sv2} at $n=1$.}\label{compare}
\end{figure}

Comparing the accuracy of both methods for $d=2$ in $10$th order, we find
that the effective potential result has an error of about $2.4\%$, 
while the strong-coupling result turns out to have an error of  about $5.0\%$. Thus we conclude 
that HEPLT is more accurate than HSSCE up to the same order. From Fig.~\ref{compare} we also read off that the HEPLT result always gives smaller hopping values than the QMC result, while the
HSSCE result gives larger values.  Thus,
we could use both methods in order to determine the region in the quantum phase diagram, where the
phase boundary must exist. Extrapolating the results for both methods to infinite order yields quantum phase boundaries, which are basically indistinguishable from QMC \cite{sv2}
in the $2$d system. However in $3$d a comparison with results from QMC \cite{sv1} reveals that extrapolating the $10$th order of HSSCE
has the same accuracy as extrapolating the $8$th order of HEPLT.\cite{martin} Thus, up to the same order, the HEPLT
method turns out to be more accurate in higher dimensional systems. 

In a previous paper \cite{tao1} we have pointed out the intriguing observation that, up to the second order, the SC coefficients coincide with those of a hopping expansion of the EPLT quantum phase boundary.
Thus, for the Mott lobe $n$, one obtains the SC upper (lower)
critical line by performing a hopping expansion of the EPLT quantum phase boundary around $\mu/U=n-1$ ($\mu/U=n)$.
In view of proving such relations also for higher orders, we perform a symbolic calculation for the HEPLT method up to the $8$th order. By a corresponding hopping expansion of the HEPLT quantum
phase boundary we obtain, indeed, the same coefficients of both the upper and the lower HSSCE critical lines. As shown in Fig.~\ref{compare}, the QMC results are located between the HSSCE and the HEPLT
critical lines, so both methods can be considered as an overestimation and an underestimation, respectively. We think the reason is that, HSSCE neglects high order corrections but HEPLT includes more.

\section{conclusion}\label{sec_con}

In this paper we developed a symbolic implementation 
the process-chain approach \cite{eckardt,martin} for applying the strong-coupling expansion method \cite{Monien} to very high orders.  The resulting analytic symbolic 
expressions allow 
a detailed analysis of how the critical lines depend on the fillings numbers and the dimensions. As a concrete example we have calculated for the Bose-Hubbard model the quantum phase diagram between the superfluid and the
Mott insulator in different dimensions. Applying the Pad{\'e} resummation method, we have obtained the critical lines in the thermodynamic limit for arbitrary fillings. A comparison with DMRG and QMC calculations the phase transition at filling $n=1$ 
demonstrates that the HSSCE method is quite accurate, so that 
it can be also used reliably determine
the transition lobes at higher fillings which are hard to obtain from 
recent numerical and analytic methods.

In addition, the analytic expression of the phase boundaries can help to investigate systematically the rescaling properties of the quantum phase diagram. At the lobe tips we have found that the rescaling 
functions of the critical hopping almost coincide for both two and three dimensions, while the corresponding rescaling functions of the critical chemical potential turn out to be different. With this rescaling
at the lobe tip
it has then been possible to approximately map all the Mott lobes of different fillings to the infinite filling Mott lobe.

Finally, we have compared the HEPLT with the HSSCE method developed here and have concluded that the latter is easier to implement. However, it has also turned out that HSSCE is less accurate as HEPLT up to the same order.
Furthermore, we have verified an intimate relation been both methods. Namely, the quantum phase boundaries from HEPLT and HSSCE have turned out to agree up to the $8$th order in a power series with respect to the hopping.

In addition, we present our algorithm in greater 
detail in Appendix A, so that the respective steps should be reproducible for the reader. Further information is accessible
in a Matlab code in the Supplemental Material, which can be downloaded from our homepage.\cite{sm}

We conclude that our paper works out in detail a general symbolic high-order perturbation theory,
whose applicability is not limited to the Bose-Hubbard model. Instead it may also be suitable to analyze other lattice systems, which describe, for instance, the supersolid-solid transition,\cite{ss1}
        a mixture of two bosonic species,\cite{sc2,sc2b} three-body interactions,\cite{zhang4} and even frustrated systems like Kagome superlattices \cite{kagomelattice} suffering from the sign problem.
        Furthermore it should be noted that our theoretical high-precision results could, in principle, be checked with in situ density measurements \cite{density} and
        single atom detection \cite{singleatom}
        as they are possible these days, for instance,  with the quantum gas microscope \cite{microscope} or the scanning electron microscope.\cite{electron}
%
\section*{Acknowledgments} 
We are thankful for useful discussions with A. Eckardt, C. Heil, and M. Holthaus. This work was supported by the "Allianz f\"{u}r
Hochleistungsrechnen Rheinland-Pfalz", by the German Research
Foundation (DFG) via the Collaborative Research
Centers SFB/TR49 and SFB/TR185, as well as by the Chinese Academy of Sciences via the Open Project Program of the State Key Laboratory of
Theoretical Physics.  This work was also supported by the
Startup Foundation for Introducing Talent of NUIST, No. $2015r060$ and
the Special Foundation for theoretical physics Research Program of
China, No 11647165.  X.-F. Z. acknowledges
funding from Project No. 2018CDQYWL0047 supported by the Fundamental Research
Funds for the Central Universities and from the National Science Foundation of
China under Grants No. 11804034 and No. 11874094.

\begin{appendix}
\section*{Appendix A: Algorithm of high-order strong-coupling expansion}\label{sec_ag}
\label{AA}
When a strong-coupling perturbation theory is used to determine the ground-state energy of a Bose-Hubbard model,
the respective orders can be calculated recursively. To this end one has to consider
different types of hopping processes
which contribute to the considered perturbative order of the ground-state energy. However, such
an algorithm has two disadvantages. (a) High-order results are based on
lower orders. (b) It is hard to automatically generate the relevant hopping processes.
In order to overcome those problems, Eckardt suggested to use the Kato
representation of 
perturbation theory.\cite{eckardt} This allows to produce the respective perturbative terms in each 
order via a process chain approach which generates and evaluates the respective diagrams systematically.
In view of the high-order strong-coupling expansion, we also use similar strategies. In the
following, we discuss the three parts how to implement the
algorithm in detail and take the lower-order terms for the
Bose-Hubbard model in a square lattice as illustrative examples.

\subsection*{A.1 Kato representation}\label{katolist}
%
%
Kato worked out a particular representation for the perturbative
terms of 
perturbation theory.\cite{kato}
Therein, the $n$th order contribution to the ground-state
energy for a perturbation of the Hamiltonian $H'$ is given by the trace
\begin{equation}
E^{(m)}={\rm Tr}\sum_{\{\alpha_{\ell}\}}
S^{\alpha_{1}}H'S^{\alpha_{2}}H'...H'S^{\alpha_{m+1}} \, .
\label{trace}
\end{equation}
Here each term is characterized by a {\it Kato trace list}
\begin{equation}
\langle \alpha_{1} \alpha_{2}... \alpha_{m+1} \rangle \, ,
\end{equation}
where the integers $\alpha_{1}, \ldots , \alpha_{m+1}$ fulfill the condition
\begin{equation}
\sum_{\ell=1}^{m+1}\alpha_{\ell}=m-1, \hspace*{0.5cm} \alpha_{\ell}\ge 0\label{sum} \,.
\end{equation}
Furthermore, the operators $S^{\alpha_{\ell}}$ are defined via
\begin{equation}
S^{\alpha_{\ell}}=
\begin{cases}
-|g\rangle \langle g| & \mbox{if $\alpha_{\ell}=0$}\\
\frac{|e\rangle \langle
	e|}{\left(E_{g}^{0}-E_{e}^{0}\right)^{\alpha_{\ell}}} &\mbox{if
	$\alpha_{\ell} \neq 0$} \, ,
\end{cases}
\label{salpha}
\end{equation}
where $|g\rangle, |e\rangle$ and $E_{g}^{0}, E_{e}^{0}$ denote wave functions and energies of
the ground state and the excited states, respectively. Thus, we read off from Eq.~(\ref{salpha}) the relation
\begin{equation}
S^{\alpha_{i}}S^{\alpha_{j}}=
\begin{cases}
-S^{0} &\mbox{$\alpha_{i}=\alpha_{j}=0$} \\
0 &\mbox{$\alpha_{i}=0,\alpha_{j} \neq 0$ or $\alpha_{i} \neq 0,\alpha_{j}=0$}\\
S^{\alpha_{i}+\alpha_{j}} &\mbox{$\alpha_{i} \neq 0,\alpha_{j}
	\neq 0$}
\end{cases}
\label{K_con}
\end{equation}
which is useful for simplifying products of operators $S^{\alpha_{\ell}}$.
Additionally, we can conclude from Eq.~(\ref{sum}) that there are at least two numbers
in the Kato trace list $\langle \alpha_{1} \alpha_{2}... \alpha_{m+1} \rangle$ which are zero.
The goal is to use Eq.~(\ref{K_con}) and the cyclic permutation of operators under the trace
to move the $S^0$-operators to the outside of the trace, so the expression can be rewritten
in general as
\begin{equation}
\langle g|H'S^{\alpha_{1}'}H'...S^{\alpha_{m-1}'}H'|g \rangle. \label{katolist0}
\end{equation}
If $\alpha_1 = \alpha_{m+1}= 0$ we immediately obtain the new indices 
$\alpha_j'=\alpha_{j+1}$.
If $\alpha_1\neq \alpha_{m+1}=0$ or $\alpha_1\neq \alpha_{m+1}=0$ the expression vanishes 
according to Eq.~(\ref{K_con}), so this case does not need to be considered.  
Finally, if $\alpha_1 \neq 0$ and  $\alpha_{m+1} \neq 0$
cyclic permutations are used until $S^0=-S^0S^0$ appears first in the product under 
the trace, which can then again be written 
in the form of Eq.~(\ref{katolist0}) with a negative
sign and one of the new indices with value $\alpha_1 + \alpha_{m+1}$.
We define the resulting reduced list of indices as a {\it Kato-list} 
and use the abbreviated representation
\begin{equation}
\label{defkatolist}
\left(\alpha_{1}' \alpha_{2}'... \alpha_{m-1}' \right).
\end{equation}
%
%
So far two ways have been used to calculate the Kato-lists.
One was suggested by Eckardt \cite{eckardt} and can be realized by the following steps.
At first we substitute the operator $S^0$ with $-|g\rangle\langle g|$ into the
Kato-list and use the abbreviation $|g\rangle\rightarrow)$ and
$\langle g|\rightarrow($. With this the Kato-list changes into an array of
elementary matrix element (EME) denoted by $(.....)$, in which the operator
$S^0$ no longer exists. Note that the EME has the reflection symmetry, so we have, for instance,
$(\alpha_{1}\alpha_{2}\alpha_{3})=(\alpha_{3}\alpha_{2}\alpha_{1})$. For convenience reasons
we always change the form of the EME such that we take the smaller one,
e.g.~we use $(121321)$ and not $(123121)$. After that, to order the array of
EMEs, we define the relative value of the EME with the following 
two rules. (i) The numbers $\alpha$ of the operators $S^{\alpha}$ in the EME are firstly
compared, thus we have $(111)>(12)$. (ii) When the numbers coincide, then we
compare the integer $\alpha$ of the first non-equal operator $S^{\alpha}$, so we have
$(123)<(132)$. Along these lines we can order the EMEs in the
Kato-list. Consider, for instance, 
$\langle 20210{\bf 1}1001\rangle$ as an example for a Kato trace list, which
can be transformed into the Kato-list $-({\bf 1}1003021)$.\cite{eckardt} 
The resulting EMEs $(11)()(3)(21)$ can then be ordered to obtain 
$()(3)(11)(12)$, which is stored as a Kato-list $-(03011012)$.
At the end, the final Kato-lists can be ordered according to similar rules
as the array of EMEs.

%
%
Let us take the fourth-order perturbative term as a concrete example to show how
to generate the Kato-lists step by step and how to order them for later
usage. It can be proved that the resulting fourth-order Kato representation coincides
with the standard result of 
perturbation theory.
\begin{widetext}
	\begin{center}
		\begin{tabular}{p{7.2cm} p{.5cm}| p{9.5cm}}
			\multicolumn{1}{c}{Algorithm} & & \multicolumn{1}{|c} {Output} \\
			\hline
			1. Generate all Kato trace lists for the considered order $n=4$. & & $\langle 30000 \rangle$,$\langle 21000 \rangle$,$\langle 20100 \rangle$,...,$\langle 10101 \rangle$,...,$\langle 00003 \rangle$\\
			2. Neglect all terms which have a zero at one end and are non-zero at the other end due to the second line of Eq.~(\ref{K_con}). & &
			$\langle 03000 \rangle$,$\langle 02100 \rangle$,$\langle 02010 \rangle$,$\langle 00210 \rangle$,$\langle 01110 \rangle$,$\langle 00120 \rangle$,
			$\langle 01020 \rangle$,$\langle 01200 \rangle$,$\langle 00030 \rangle$,$\langle 10011 \rangle$,$\langle 10101 \rangle$,$\langle 11001 \rangle$,
			$\langle 20001 \rangle$,$\langle 10002 \rangle$,$\langle 00300 \rangle$\\
			3. Change Kato trace list to Kato-list. &
			&$(300)$,$(210)$,$(201)$,$(021)$,$(111)$,$(012)$,$(102)$,$(120)$,
			$(003)$,$-(012)$,$-(102)$, $-(210)$,$-(300)$,$-(300)$,$(030)$\\
			4. Perform the substitutions $|g\rangle\rightarrow)$ and $\langle g|\rightarrow($ and order array of EMEs. &
			&$()()(3)$,$-()(12)$,$-(1)(2)$,$-()(12)$,$(111)$,$-()(12)$,$-(1)(2)$,$-()(12)$,
			$()()(3)$,$()(12)$,$(1)(2)$,$()(12)$,$-()()(3)$,$-()()(3)$,$()()(3)$\\
			5. Collect same arrays and determine their weight. & &$-(1)(2)$,$-2()(12)$,$()()(3)$,$(111)$\\
			6. Order and change them back to Kato-list. &
			&$(003)$,$2(012)$,$(102)$,$(111)$ \\\hline
		\end{tabular}
	\end{center}
\end{widetext}
The advantage of this algorithm is that one can get for each order the smallest number of 
Kato-lists, so that the computation time is drastically reduced.
But the Kato-lists following from the above algorithm are only suitable provided that the ground state is non-degenerate. Let us illustrate this by 
the process $\langle g_{1}|H'S^{\alpha_{1}'}H'|g_{2} \rangle\langle g_{2}|H'S^{\alpha_{2}'}H'|g_{3} \rangle$,
which involves the three degenerate ground states $|g_{1} \rangle, |g_{2} \rangle, |g_{3} \rangle$ and
can be represented by the Kato-list $(\alpha_{1}'0\alpha_{2}')$. If we transform the Kato-list into EMEs and change their order, the calculation process may be changed to
$\langle g_{2}|H'S^{\alpha_{1}'}H'|g_{3} \rangle\langle g_{1}|H'S^{\alpha_{2}'}H'|g_{2} \rangle$.
But this cannot be mapped back to the Kato-list $(\alpha_{2}'0\alpha_{1}')$ as the number $0$ in the Kato-list corresponds to
$\sum_j|g_{j} \rangle\langle g_{j}|$, but not $\sum_{i,j}|g_{i} \rangle\langle g_{j}|$.

A second approach for calculating Kato-lists was proposed in Ref.~[\onlinecite{Heil}]
in order to also treat degenerate ground states, 
which was then applied to a 1D system.
Here the Kato-lists (\ref{defkatolist}) have to fulfill the conditions
\begin{equation}
\mathop{\sum}_{\ell=1}^{m-1}\alpha_{\ell}'=m-1, \hspace*{0.5cm}\alpha_{\ell}'\ge 0
\label{katosum}
\end{equation}
and
\begin{equation}
\sum_{\ell=1}^{s}\alpha_{\ell}'\leq s, \hspace*{0.5cm} \mbox{for $s=1,...,m-2$}\,.
\label{katosum2}
\end{equation}
Thus, more Kato-lists appear in each order, so the computational effort increases. In particular, now each Kato-list appears with the multiplicity one.
For instance, in fourth order we no longer have the Kato-lists $(003)$,$2(012)$,$(102)$,$(111)$ for a non-degenerate ground state but instead
$(003), (012), (021), (102), (111)$.

It should be remarked that in the algorithm the generated Kato-lists will be stored in binary form using a suitable hashing technique, so that it is efficient to search for matching Kato-lists, which will be useful for the calculation below.

In conclusion, the original Kato-lists in Ref.~[\onlinecite{eckardt}]
are useful for calculating non-degenerate ground-state energies as they show up, for instance, for Mott states.
In contrast to this, the degenerate states of particle and 
hole excitations, as they appear within the strong-coupling method, have to be determined
from Kato-lists of multiplicity one by taking into account
the restrictions in Eqs.~(\ref{katosum}) and (\ref{katosum2}).\cite{Heil}

\begin{figure}[t]
	\includegraphics[width=0.5 \textwidth]{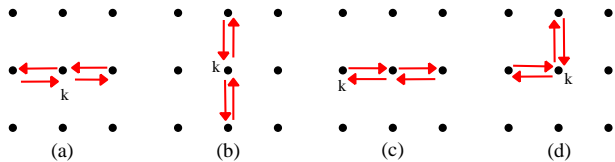}
	\caption{Arrow diagrams of 4th order in the square lattice, which are topologically equivalent}
	\label{repeated}
\end{figure}
\subsection*{A.2 Arrow diagrams generalization}
%
%
Whereas the determination of the underlying Kato-lists is independent of the considered system,
we now turn to their diagrammatic representation, which does depend on the underlying Hamiltonian
or the topology of the lattice. To this end we follow Refs.~\cite{eckardt,Heil} and remark that 
an expression $\langle e_1|H'|e_2\rangle$ in the Kato-list 
corresponds to a hopping process on the lattice, which
can be represented by an arrow. Thus, in a lattice
system, each perturbative term can be graphically depicted as an arrow diagram.
According to the linked cluster theorem, only connected diagrams contribute to the ground-state energy, thus we
only need all non-equivalent connected arrow diagrams.
Whereas for the calculation of the non-degenerate ground-state energy only
closed connected diagrams appear, for degenerate ground-state energies
also open connected diagrams have to be considered.
For instance, for a particle (hole) excited degenerate ground state, each perturbative term is equivalent to the hopping of a particle (hole)
to another site, so any open connected diagram has exactly two ends. Thus, 
the respective arrow diagram can be interpreted as the path of a moving particle.

%
%
In order to generate all non-equivalent arrow diagrams, we fix at first
the starting point at the center, and use different numbers or
characters to label the respective directions. For instance, in case of a square
lattice, we abbreviate up as `$u$', down as `$d$', left as `$l$', and right as `$r$'.
After that, we get all possible connected arrow diagrams by applying combinatorics
and represent them with the corresponding arrays of characters. Let us consider
the second-order arrow diagrams as a concrete example: in case of a square lattice
we have in total $16$ diagrams, but only $(ud)$, $(du)$, $(lr)$, and
$(rl)$ represent closed diagrams, whereas the others are open ones.
Each diagram can have a non-unique representation, such as the 4th-order diagram in 
Fig.~\ref{repeated}(a), which can be represented by $(lrrl)$ or by $(rllr)$. But if we
define the priority $r<u<l<d$ and only take the smallest one according to this order,
the array corresponding to Fig.~\ref{repeated}(a) is identified with $(rllr)$. Proceeding in this way we avoid an overcounting of diagrams.

\begin{figure}[t]
	\vspace{-15px}
	\includegraphics[width=0.5 \textwidth]{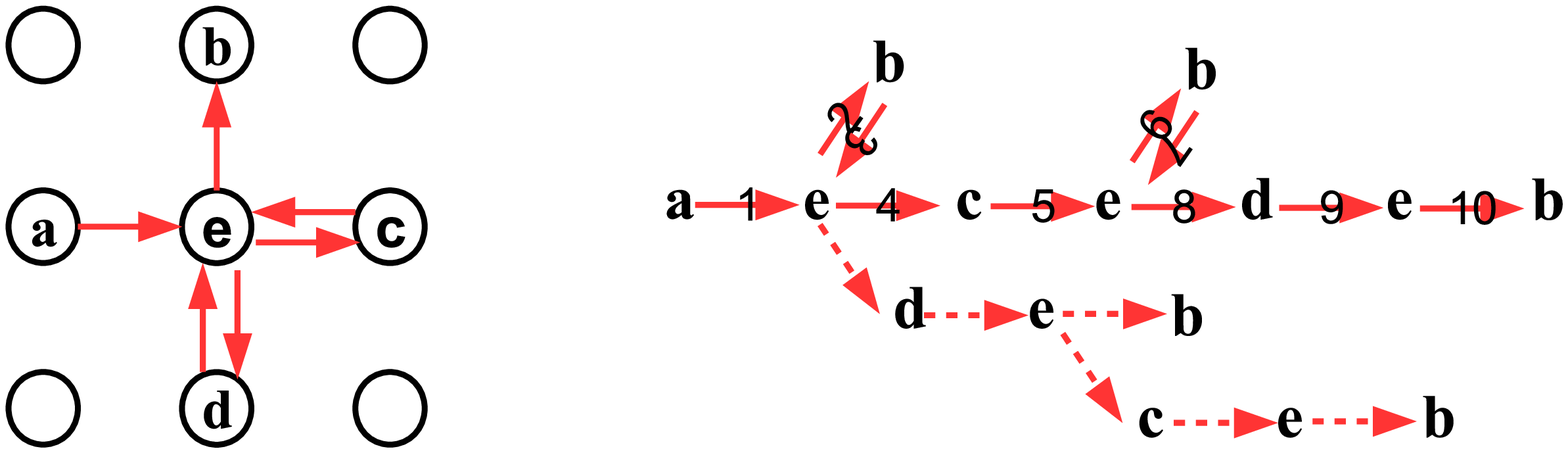}
	\vspace{-70px}
	\caption{One $6$th-order arrow diagram labeled with $(rdurlu)$ is shown in the left panel, its corresponding tree graph is shown
		in the right panel. Red solid arrows in the right panel are paths, which are passed due to the algorithm, and the dashed arrows represent other
		possible paths. The arrows $(2,3,6,7)$ indicate trace back process, so the smallest path is $(1,4,5,8,9,10)$. Thus, the smallest
		diagram representation is provided by $(rrlduu)$.}\label{overcounting}
\end{figure}

However, judging whether the diagram representation is the smallest one is not trivial, since there are several ways to follow a given set of arrows that point in and out from a central
site as shown in Fig.~\ref{overcounting}. 
All possible paths can be drawn in form of a tree graph as shown in the right panel of
Fig.~\ref{overcounting}. The smallest path, which passes through all the arrows, is the one to be determined.
To this end, 
one can go through the whole diagram by moving in each step through the smallest arrow, but 
it is not allowed to move through one and the same arrow again.
If one cannot finish with passing all the arrows, one has to trace back to the latest 
site, which
has at least two arrows pointing out, and has now to choose the next smallest arrow to move through. After several tracing
back processes, one gets the smallest diagram representation, when all arrows are passed. 

%
%
After having produced all the smallest diagrams of the considered perturbative order, we reduce the number
of diagrams due to symmetries and take the number of
equivalent diagrams as its corresponding weight. For the lattice system, we can 
take into account
both point group symmetry and translational symmetry. Taking again
the square lattice as a concrete example, its point symmetry D$_4$ has 8
operations. For instance, diagram Fig.~\ref{repeated}(b) can be obtained by
rotating Fig.~\ref{repeated}(a) by 90 degrees to the right. After implementing all 8 operations
on Fig.~\ref{repeated}, we obtain only two different diagrams (a) and
(b), so their weight is 2. Furthermore, for the Mott insulating state we can also 
consider the translational symmetry. The diagram Fig.~\ref{repeated}(c) can be obtained by moving the diagram Fig.~\ref{repeated}(a) one
lattice site to the right. As the diagram Fig.~\ref{repeated}(a) covers 3 sites,
the total number of group operations is $8\times3=24$. After having performed all these
operations on diagram Fig.~\ref{repeated}(a), we find that there are in total six diagrams with the
same contribution, so its weight is finally six. 
For the particle (hole) excited state, the starting point should always be
at site $k$, so there is no translational symmetry and only the 
point group symmetry has to be
considered. Based on group theory, we find a more generic method to
determine the weight of the smallest arrow diagram. To this end we denote the symmetry group by $G$ and
the number of its elements by $n_G$. If an arrow diagram is not changed under $n_g$ group operations,
this means that these operations belong to a subgroup $g$ of $G$ and that there are $n_G/n_g$ different arrow
diagrams. These arrow diagrams can be obtained by performing group operations on each other and they are
not changed under operations, which belong to the subgroup  $g$. The smallest arrow diagram can
be stored and its weight turns out to be $n_G/n_g$. Note that, in contrast to calculating the ground-state energy of
the Mott insulator, for particle (or hole) energy corrections, the weight does not need
to be divided by the number of covered sites, since no overcounting exists.
In Table~\ref{diagram} we show
the number $n_s$ of diagrams in the square lattice after having applied the reduction by symmetry up to 12th perturbative order. For energy corrections,
we consider only closed diagrams with both point group and translational
symmetries. But for the SC energy both closed and open
diagrams appear, but a simplification occurs only due to point group symmetries. Consequently,
although both energies contain closed diagrams,
they have different weights due to the different symmetry
considerations.

\begin{table}\begin{center}
		\begin{tabular}{|c|c|c||c|c|}
			\hline
			&\multicolumn{2}{c||}{ $~$Energy correction$~$ } &\multicolumn{2}{c|}{ SC energy } \\
			\hline
			$i$ & $n_s$ & $n_t$ & $n_s$ & $n_t$\\
			\hline
			1 & 0 & 0 & 1 & 1\\
			\hline
			2 & 1 & 1 & 3 & 2\\
			\hline
			3 & 0 & 0 & 10 & 4\\
			\hline
			4 & 4 & 3 & 36 & 10\\
			\hline
			5 & 0 & 0 & 129 & 22\\
			\hline
			6 & 12 & 7 & 477 & 58\\
			\hline
			7 & 0 & 0 & 1784 & 140\\
			\hline
			8 & 75 & 29 & 6668 & 390\\
			\hline
			9 & 0 & 0 & 24909 & 988\\
			\hline
			10 & 510 & 121 & 92748 & 2815\\
			\hline
			11 & 0 & 0 & 344907 & 7412\\
			\hline
			$~$12$~$ & $~~$4284$~~$ & 698 & $~$1278092$~$ & $~$21516$~$\\
			\hline
		\end{tabular}
		\caption{Number of different connected arrow diagrams in square lattice for Mott energy correction and
			strong-coupling energy after reduction due to symmetries ($n_s$) and topologies ($n_t$).}\label{diagram}
\end{center}\end{table}
%
%
%
Despite this symmetry simplification, we can further reduce the
diagram number by considering their topology. As there is no
long-range diagonal interaction and the on-site Hamiltonian $H_0$
is uniform, the diagrams (a) and (d) of Fig.~\ref{repeated} have
the same value. By marking the respective positions between
different sites in the diagram and relabeling the involved site indices, we can find all topological
equivalent arrow diagrams and store the smallest one at the end. Then, the sum of their
weights is considered to be the new weight of the diagram. For the diagrams shown in
Fig.~\ref{repeated}, only (c) needs to be stored.
From Table \ref{diagram}, we find that the reduction due to topology 
dramatically decreases the number of diagrams $n_t$, especially for
the open diagrams which represent the dominant part of the particle (or hole) energy.
Moreover, note that for a bipartite lattice no closed diagram exists 
in any odd perturbative order.

Thus, the algorithm concerning the diagrammatic representation of Kato-lists is summarized as follows. At first all connected
arrow diagrams are produced by combinatorics, discarding the overcounted ones.
Afterward, both symmetry and topology considerations
reduce the number of different arrow diagrams.

\begin{figure}[t]
	\includegraphics[width=0.5 \textwidth]{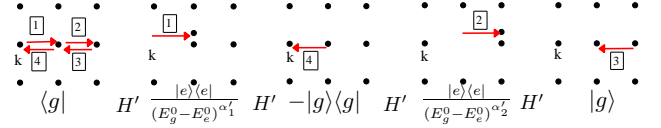}
	\caption{One process chain of energy correction Kato-list with arrow order 1 4 2 3 matching the Kato-list $(\alpha_1'0\alpha_2')$.}
	\label{pc1}
\end{figure}

\subsection*{A.3 Calculation} \label{calculation}
%
%
At last, we turn to the explicit calculation of the energy corrections for the non-degenerate Mott state.\cite{eckardt}
To this end one has to take into account that there are several ways for each arrow diagram to specify the arrow order and each order list is called a \emph{process chain}.
To illustrate this by a concrete example, we consider again the fourth-order diagram in the square lattice. 
Based on the arguments above, we only store in Fig.~\ref{repeated}
the arrow diagram (c), which is characterized by $(rrll)$. If we label each arrow with
a number, we have in total four processes. Taking into account all permutations, we have $4!=24$
types of process orders or process chains. In Fig.~\ref{pc1}, we show
one possible process chain with arrow order 1 4 2 3. All Kato-lists, which satisfy the form $(\alpha_1'0\alpha_2')$	with $\alpha_1'\ne0$,$\alpha_2'\ne0$, contribute to the corresponding energy
correction of this process. From Appendix A.1 we know that only the Katolist $(102)$ appears, which can be calculated. For the Mott insulator with filling $n$, the
energy correction of this process chain is given by
$\frac{(-t\sqrt{n}\sqrt{n+1})^4}{(-U)(-1)(-U)^2}$, which has to be multiplied with the
weight of the Kato-list $(102)$, i.e.~with one. After having obtained all the values of the
process chains of this arrow diagram, we need to sum them and multiply the result
with the weight of this diagram, which is 4.

\begin{figure}[t]
	\includegraphics[width=0.5 \textwidth]{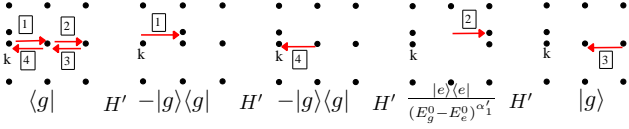}
	\caption{One process chain of SC energy Kato-list with arrow order 1 4 2 3 matches Kato-list $(0 0\alpha_1')$.}\label{pc2}
\end{figure}

%
%
For calculating the SC energy, we need to consider the degenerate
ground state and the algorithm has to be adjusted accordingly.\cite{Heil} The main idea for
calculating the degenerate ground state energy is to construct the
effective Hamiltonian matrix perturbatively in the Hilbert space of the degenerate ground states.
The diagrams, which have the same end points, belong to the same matrix elements and 
define an effective hopping process.\cite{ss1}
Thus in order to get the SC energy, we only need to add all effective 
hopping matrix elements.
Similar to the Mott energy correction, we first need to find out all the process chains and their suitable
Kato-lists. As shown in the process chain of Fig.~\ref{pc2}, any
state $\hat{b}_i^{\dag} |\psi_0\rangle$ represents a zeroth order ground state, so the
process chain matches the Kato-list $(00\alpha_1')$, and
the only Kato-list of this form, which appears, is $(003)$. In case of a particle excitation of a Mott insulator with filling $n$, the energy correction
of this process chain turns out to be $\frac{[-t(n+1)]^2[-t \sqrt{n(n+1)}]^2}{(-1)(-1)(-U)^3}$. After having obtained all the values of the
process chains of this arrow diagram, we need to sum them and multiply the result
with the weight of this diagram, which is 6.

\subsection*{A.4 Summary}

In conclusion, we have to proceed along the following lines in order to calculate the SC
energy: At first,  we have to generate all process chains of the arrow diagram. Then we have to
find the suitable Kato-lists, which we have to calculate. We sum 
all energy values of the different Kato-lists, in order to get the energy of
one process chain. Subsequently, we sum the energies of all process chains and
multiply with the weights of the arrow diagrams to determine the energy of
the arrow diagrams. At last, we sum the energies of all open and closed
arrow diagrams for the SC energy. Finally, by subtracting the energy correction of the particle (or hole) excitation and the Mott state, we obtain the resulting critical
line in the respective perturbative order. 

\section*{Appendix B: : Strong-coupling quantum phase boundary for arbitrary Mott lobe}

\begin{widetext}
Here we present our strong-coupling results for the critical lines in Eqs.~(\ref{particle}) and (\ref{hole}) of the Bose-Hubbard quantum
phase diagram for a general Mott lobe $n$, which were obtained with symbolic calculations up to the 8th order. In order to simplify our notation we
give in this appendix both the chemical
potential and the hopping matrix element in terms of the energy unit scale of $U$.
At first,
we consider the Bose-Hubbard chain, i.e. the case $d = 1$, where the upper boundary reads
\begin{align}
\mu_{p}=&n-2(n+1)\frac{t}{U}+n^2 t^2
  +\left(n^3+3 n^2+2 n\right) t^3+\frac{1}{120} \left(221 n^4-70 n^3-559 n^2-270 n\right) t^4+\frac{1}{3600}\left(11063 n^5 \right. \nonumber \\ \nonumber
&\left.+56453 n^4+127443 n^3+114703 n^2+32650 n\right) t^5+\frac{1}{9072000}\left(4201738 n^6-159047706 n^5-1019369927 n^4 \right. \\ \nonumber
&\left.-1684710576 n^3-1075545275 n^2-246900750 n\right) t^6+\frac{1}{11430720000}\left(65154949955 n^7+852816554717 n^6 \right.\\ \nonumber
&\left.+4617015079493 n^5+10042635987533 n^4+10398278064548 n^3 +5217933284246 n^2+1032937732500 n\right) t^7\\ \nonumber
&-\frac{1}{35012981203200000}\left(200810348470291303 n^8+5778066670166672114 n^7 +43424519219943077927 n^6\right. \\ \nonumber
&+134818712387825712764 n^5+207253111515652899680 n^4+168130666829189484230 n^3 \\ 
&\left.+69538136222775683026 n^2+11689091525838787500 n\right) t^8\,, \label{1dup}
\end{align}
whereas the lower boundary is given by
\begin{align}
\mu_{h}=& n-1 +2 n t -\left(n^2+2 n+1\right) t^2+\left(-n^3+n\right) t^3
  +\frac{1}{120} \left(-221 n^4-954 n^3-977 n^2-246 n-2\right) t^4\nonumber \\ \nonumber
&+\frac{1}{3600}\left(-11063 n^5+1138n^4-12261 n^3-39538 n^2-15076 n\right) t^5+\frac{1}{9072000}\left(-4201738 n^6-184258134 n^5 \right. \\ \nonumber
&\left.+161105327 n^4+718257312 n^3+484129979 n^2+107088822 n+54432\right) t^6+\frac{1}{11430720000}\left(-65154949955 n^7 \right. \\ \nonumber
&\left.+396731905032 n^6-868369700246 n^5-2530614337602 n^4-1621977063431 n^3-467241407430 n^2 \right.\\ \nonumber
&\left.-45622126368 n\right) t^7+\frac{1}{35012981203200000}\left(200810348470291303 n^8-4171583882404341690 n^7 \right. \\ \nonumber
&+8600742285944529613 n^6+15634382372668953372 n^5-3648271187242628015 n^4-9801914187470665632 n^3 \\ 
&\left.-3871241066318288229 n^2-378804028118746050 n+39893821295328\right) t^8\,. \label{1ddown}
\end{align}
Correspondingly, the two-dimensional lattice system turns out to have the upper boundary 
\begin{align}
\mu_{p}=&n-4(n+1) t-2(3n^2+4n) t^2-4\left(11 n^3+15 n^2+4n\right) t^3
  -\frac{1}{60}\left(10597 n^4+24490 n^3+17857 n^2+4050n\right) t^4 \nonumber \\ \nonumber
&-\frac{1}{450}\left(941863 n^5+2188403 n^4+1638443 n^3+387253n^2-4650n\right) t^5-\frac{1}{1512000} \left(17584506524 n^6 \right.\\\nonumber
&\left.+56792815090 n^5+70292096771 n^4+41746636380 n^3+12245886137 n^2+1586224250n\right) t^6-\frac{1}{952560000}  \\\nonumber
&\left(143621438541217 n^7+482685930471859 n^6+614916508856545 n^5+357580517960125 n^4 \right. \\\nonumber
&\left.+81552627354718 n^3-3273334379504 n^2-3097460700000n \right) t^7 -\frac{1}{17506490601600000}\\\nonumber
& \left(15828728653296464701673 n^8+66223539683880680581462 n^7+114520168944419866201057 n^6 \right.\\\nonumber
&+106687922443046182561852 n^5+59150401403670755097004 n^4+20678487502333510720450 n^3 \\
&\left.+4647036608792957074250 n^2+556471872345686887500n\right) t^8\,, \label{2dup}
\end{align}
and the lower boundary is determined by
\begin{align}
\mu_{h}=&n-1+4 n t+\left(6 n^2+4 n-2\right) t^2+4 n (n+1) (11 n+7) t^3+\frac{1}{60} \left(10597 n^4+17898 n^3+7969 n^2+582 n \right.\nonumber\\ \nonumber
&\left.-86\right) t^4+\frac{1}{450}\left(941863 n^5+2520912 n^4+2303461 n^3+816288 n^2+91876 n\right) t^5+\frac{1}{1512000}\left(17584506524 n^6 \right. \\\nonumber
&\left.+48714224054 n^5+50095619181 n^4+23183730284 n^3+4598004583 n^2+376989662 n-3186288\right) t^6 \\\nonumber
&+\frac{1}{952560000} \left(143621438541217 n^7+522664139316660
n^6+734851135390948 n^5+503463418187310 n^4 \right. \\\nonumber
&\left.+173427383585083 n^3+27374449536030 n^2+1602049522752n\right) t^7+\frac{1}{17506490601600000}\\\nonumber
&\left(15828728653296464701673 n^8+60406289542491037031922 n^7+94159793449556113777667 n^6 \right.\\\nonumber
&+76147562446580745727476 n^5+33700440149666544069539 n^4+8032188219063211012704 n^3 \\
&\left. +947361190459842328401 n^2+50197343417953427898 n-85891426017677280\right) t^8\,. \label{2ddown}
\end{align}
Finally, the upper boundary of a three-dimensional Mott lobe reads
\begin{align}
\mu_{p}=&n+(-6n-6)t-3 \left(7 n^2+8 n\right)t^2+\left(-231 n^3-333 n^2-102 n\right)t^3+\frac{1}{8} \left(-20683 n^4-42646 n^3-26919 n^2 \right.\nonumber\\\nonumber
&\left.-4990 n\right) t^4+\frac{1}{240} \left(-9450991 n^5-23212861 n^4-19367531 n^3-6129431 n^2-523770 n\right) t^5 \\\nonumber
&+\frac{1}{3024000}\left(-1612512639802 n^6-4903150529490 n^5-5568888652747 n^4-2895859756680 n^3-673691826835 n^2 \right.\\\nonumber
&\left.-56132922750 n\right) t^6+\frac{1}{1270080000}\left(-11585180188993081 n^7-40169594556544231 n^6-54166855276590991 n^5 \right.\\\nonumber
&\left.-35546615871621511 n^4-11500224469294300 n^3-1582089821730130 n^2-46040315017500 n\right)t^7\\\nonumber &+\frac{1}{11670993734400000}\left(-1591704030486631948004941 n^8-6411307825175115072269186 n^7\right.\\\nonumber
&-10477219584627901596466499 n^6-8903631755605628211005096 n^5-4192954829256834761500718 n^4\\
&\left.-1081820439715328380764170 n^3-143411169994732620393730 n^2-8530598518867251187500 n\right) t^8\,, \label{3dup}
\end{align}
where the low boundary is characterized by
\begin{align}
\mu_{h}=& n-1 +6nt+3 \left(7 n^2+6 n-1\right) t^2+\left(231 n^3+360 n^2+129 n\right) t^3+\frac{1}{8} \left(20683 n^4+40086 n^3+23079 n^2 \right.\nonumber\\\nonumber
&\left.+3642n-34\right) t^4+\frac{1}{240} \left(9450991 n^5+24042094 n^4+21025997 n^3+7205906 n^2+771012 n\right) t^5+\frac{1}{3024000}\\\nonumber
&\left(1612512639802 n^6+4771925309322 n^5+5240825602327 n^4+2598442355448 n^3+555628775407 n^2 \right.\\\nonumber
&\left.+38549263230 n-50089536\right) t^6+\frac{1}{1270080000}\left(11585180188993081 n^7+40926666766407336 n^6 \right.\\\nonumber
&+56438071906180306 n^5+38225048777927814 n^4+13071729232591381 n^3+2052306743024850 n^2 \\\nonumber
&\left.+109040959595232 n\right)t^7+\frac{1}{11670993734400000}\left(1591704030486631948004941 n^8
\right.\\\nonumber
&+6322324418717940511770342 n^7+10165777662027790634720545 n^6+8457647130735753938417688 n^5 \\\nonumber
&+3856598073582426484397083 n^4+937724839936122043372908 n^3+107903564700275910537495 n^2\\
&\left.+4285936758470495239062 n-1004648837988860064\right)t^8\,. \label{3ddown}
\end{align}	
\end{widetext}

\end{appendix}

\end{document}